\documentclass[aps,prd,preprint,superscriptaddress,tightenlines,nofootinbib]{revtex4}

\usepackage{amsfonts,amsmath,amsthm,amssymb}
\usepackage{mathrsfs}
\usepackage{bm}
\usepackage{color}
\usepackage{epsfig}

\usepackage[usenames,dvipsnames]{xcolor}
\definecolor{orange}{cmyk}{0,0.5,1,0}
\definecolor{rossoCP3}{cmyk}{0,.88,.77,.40}
\definecolor{graa}{rgb}{0.8,0.8,0.8}
\definecolor{blaa}{rgb}{0.2,0.2,0.6}

\newcommand{\PRE}[1]{{#1}}   % Use if preprint style
\newcommand{\met} {\not\!\! E_T}

\newcommand{\comment}[1]{}

\newcommand{\beq}[1]{\begin{equation}\label{#1}}
\newcommand{\eeq}{\end{equation}}
\newcommand{\be}{\begin{equation}}
\newcommand{\ee}{\end{equation}}
\newcommand{\beqa}[1]{\begin{eqnarray} \label{#1}}
\newcommand{\eeqa}{\end{eqnarray}}
\newcommand{\bea}{\begin{eqnarray}}
\newcommand{\eea}{\end{eqnarray}}
\newcommand{\bay}[1]{\left(\begin{array}{#1}}
\newcommand{\eay}{\end{array}\right)}
\newcommand{\ba}{\begin{array}}
\newcommand{\ea}{\end{array}}
\newcommand{\rf}[1]{(\ref{#1})}

\newcommand{\gs}{\mathrel{
   \rlap{\raise 0.511ex \hbox{$>$}}{\lower 0.511ex \hbox{$\sim$}}}}
\newcommand{\ls}{\mathrel{
   \rlap{\raise 0.511ex \hbox{$<$}}{\lower 0.511ex \hbox{$\sim$}}}}

\newcommand{\rarr}{\rightarrow}

\newcommand{\Emin}{E_\nu^{\min} }
\newcommand{\Emax}{E_\nu^{\max} }
\newcommand{\ymean}{\langle y \rangle}

\newcommand{\nubar}{\bar\nu}
\newcommand{\nue}{{\nu_{e}}}
\newcommand{\numu}{{\nu_{\mu}}}
\newcommand{\nutau}{{\nu_{\tau}}}
\newcommand{\nuebar}{{\bar \nu}_e}
\newcommand{\numubar}{{\bar \nu}_\mu}
\newcommand{\nutaubar}{{\bar \nu}_\tau}

\newcommand{\nua}{\nu_\alpha}

\newcommand{\nub}{\nu_\beta}

\def\met{\mbox{${\hbox{$E$\kern-0.6em\lower-.1ex\hbox{/}}}_T$}}   %missing ET

\def\CR{{\cal R}}

\newcommand{\Nexp}{ {\langle N^{\rm expected}_{1\mbox{-} 2\,{\rm PeV}} \rangle} } 
\newcommand{\nmean}{\langle  n \rangle}

\begin{document}

\title{ \PRE{\vspace*{0.9in}}   \color{rossoCP3}
{
Glashow resonance as a window into cosmic neutrino sources
}
	
\PRE{\vspace*{0.1in}} }

%AUTHORS

% \author{L.A.~Anchordoqui}
% \affiliation{Department of Physics,\\
% University of Wisconsin-Milwaukee,
% Milwaukee, WI 53201, USA
% \PRE{\vspace*{.05in}}
%}

\author{V.~Barger}
\affiliation{Department of Physics,\\
University of Wisconsin, Madison, WI 53706, USA
\PRE{\vspace*{.1in}}
}

\author{Lingjun~Fu}
\affiliation{Department of Physics \& Astronomy,\\
Vanderbilt University, Nashville, TN 37235, USA
\PRE{\vspace*{.1in}}
}

% \author{H.~Goldberg}
% \affiliation{Department of Physics,\\
% Northeastern University, Boston, MA 02115, USA
% \PRE{\vspace*{.1in}}
% }

\author{J.G.~Learned}
\affiliation{Department of Physics \& Astronomy,\\
University of Hawaii at Manoa, Honolulu, HI 96822, USA
\PRE{\vspace*{.1in}}
}

\author{D.~Marfatia}
\affiliation{Department of Physics \& Astronomy,\\
University of Hawaii at Manoa, Honolulu, HI 96822, USA
\PRE{\vspace*{.1in}}
}

\author{S.~Pakvasa}
\affiliation{Department of Physics \& Astronomy,\\
University of Hawaii at Manoa, Honolulu, HI 96822, USA
\PRE{\vspace*{.1in}}
}

\author{T.J.~Weiler$^2$}
%\affiliation{Department of Physics \& Astronomy,\\
%Vanderbilt University, Nashville TN 37235, USA
%\PRE{\vspace*{.1in}}
%}

%\date{\today}

\begin{abstract}
 \PRE{\vspace*{.1in}} 
\noindent 
The Glashow resonance at $E_\nu =6.3$~PeV is a measure of the $\nuebar$ content of the astrophysical neutrino flux.
The fractional $\nuebar$ content depends on the neutrino production model at the cosmic neutrino source, 
and the environment at the source.  
Thus, the strength of the Glashow resonance event rate is a potential window into astrophysical sources.
We quantify the ``Glashow resonometer" and comment on the significance that no Glashow events are observed in the IceCube three--year data.
 
%Implications of this hypothesis for other physics, beginning with the
%end of hope for UHE neutrino astronomy, can hardly be
%overstated. 
%Testable repercussions are outlined.  
%
\end{abstract}
\pacs{xxxx}

\maketitle

\section{Introduction}
\label{sec:intro}
The rate of interaction of $\nu_e$, $\nu_\mu$, $\nu_\tau$, $\bar
\nu_\mu$, $\bar \nu_\tau$, with electrons is mostly negligible compared to
interactions with nucleons. However, the case of $\bar \nu_e$ is
unique because of resonant scattering, $\bar \nu_e e^- \to W^- \to 
{\rm anything}$, at $E_\nu \simeq 6.3~{\rm PeV}$.
The $W^-$ resonance in this process is commonly referred
to as the Glashow resonance~\cite{Glashow:1960zz}.  The signal for
$\bar \nu_e$ at the Glashow resonance, when normalized to the total
$\nu + \bar \nu$ flux, can be used to differentiate among the main 
primary mechanisms for neutrino-producing interactions in 
optically thin sources of cosmic rays~\cite{Anchordoqui:2004eb}. 
% For example, in the $pp$ collision mechanism, a nearly
% isotopically neutral mix of pions results, which will create on decay a neutrino
% population with the ratio $N_{\nu_\mu}=N_{\bar\nu_\mu} =
% 2N_{\nu_e}=2N_{\bar\nu_e}.$ 
% On the other hand, in photopion interactions, the isotopically asymmetric process 
% $p\gamma\rightarrow\Delta^+\rightarrow \pi^+ n$, 
% $\pi^+\to\mu^+ \nu_\mu\to e^+\nu_e \nubar_\mu \nu_\mu$,
% is the dominant source of neutrinos so that at production, 
% \mbox{$N_{\nu_\mu}=N_{\bar\nu_\mu} = N_{\nu_e} \gg N_{\bar\nu_e}$.}

In 2012, IceCube released the first two-year equivalent dataset,
observing high-energy non-atmospheric
neutrino events for the first time~\cite{Aartsen:2013bka,Aartsen:2013jdh}.
The maximum neutrino energy inferred was~1--2~PeV\@.
In 2014, IceCube reported its three-year dataset~\cite{IceCube3yr}.
The maximum neutrino energy inferred to date remains at $\sim 2$~PeV.
The energy resolution on the observed events is $\sim 25$\%.
In particular, Glashow resonance events should  produce showers that are not (yet) observed.
% The effective IceCube area at the resonant energy is 
% about 40 times that of off-resonance events at a
% PeV~\cite{Anchordoqui:2004eb}, 
%
The integrated cross section of the resonance is comparable for some flavor models to 
that of the non-resonant spectrum integrated above a~PeV, 
which implies that the falling power law ($E_\nu^{-\alpha}$) of the incident neutrino spectrum
is effectively canceled and that resonant events could have been seen~\cite{EarlierResStudies}.
% The naively expected event number at IceCube for a neutrino flux on-resonance at
% $\sim 6.3$~PeV, relative to the two observed events at $\sim$~PeV, is
% $2\times 40\times 6.3^{-\alpha}= 2 \times(6.3)^{(2-\alpha)}$.  

In this Letter, we evaluate the ratio of the expected number of Glashow events at  6.3~PeV to the number of non-resonant events expected above various minimum energies ($\sim$ PeV) for six popular cosmic neutrino source models. 

%The showers due to the Glashow resonance (the $W$-boson)
% but specific to the present context of high-energy neutrino annihilation on electrons in the detector material) 
%events should populate multiple energy peaks~\cite{Kistler:2013my}.  
%The dominant one is at $\Eres=M_W^2/2m_e=6.3$~PeV, while the others occur at
%$\Evis=\Eres-E_X$, where $E_X$ is the energy in the $W$ decay which
%does not contribute to the visible shower: the hadronic decay modes
%$W\rarr q\bar{q}$ will populate the peak at 6.3~PeV, while the leptonic modes $W^-\rarr\nubar+\ell^-$ 
%will lose about   of their energy to the invisible neutrino. 
%Furthermore, the muonic mode will show just a track, and no shower at all, while in the $\tau$ mode, 
%the $\tau$ decay will produce a second invisible neutrino, leaving a visible
%shower with about 1/4 of the energy of the initial $\nuebar$.  
%Thus we expect the ratio of the lower-energy peaks at $\sim 1.6$ and 3.2~PeV, to
%the higher-energy peak at 6.3~PeV, to be $\rm{BR}(\tau):\rm{BR}(e):\rm{BR(hadrons)}\sim 1:1:6$.  
%It may be tempting to associate the observed events at 1-2~PeV with the the leptonic decay
%modes of the Glashow resonance,
% ~\cite{Barger:2012mz}, 
%but doing so leaves unexplained the issue of why the much more numerous resonance events
%at 6.3~PeV are not seen.

\section{Six Astrophysical Neutrino Source Models}
\label{fig:SixModels}
We consider six possible source models:\\
(i) $pp\rarr \pi^\pm$~pairs $\rarr  \nue+\nuebar+2\numu+2\numubar$, referred to as the {\it ``$\pi^\pm$ mode''}; \\
(ii) $pp\rarr \pi^\pm$~pairs $\rarr \numu,\ \numubar$ only, referred to as the {\it ``damped $\mu^\pm$ mode''}; \\
(iii) $p\gamma\rarr\pi^+$ only, $\rarr \nue+\numu+\numubar$, referred to as the {\it ``$\pi^+$ mode''}; \\
(iv) $p\gamma\rarr\pi^+ \rarr \numu$ only, referred to as the {\it ``damped $\mu^+$ mode''}; \\
(v) charm production and immediate decay to $\nue,\ \nuebar,\ \numu,\ \numubar$, referred to as the {\it ``prompt mode"}; 
and \\ (vi) $\beta$-decay of cosmic neutrons to $\nuebar$, referred to as the {\it ``neutron decay (or $\beta$ decay) mode''}. \\ 
The initial flavor content of the produced neutrinos in these six models are summarized in the second column of 
Table~\ref{table:nuebar}.  

%
%\new page
%
\begin{table}[hbt]
\caption
{Neutrino flavor ratios at source, component of $\nuebar$ in total neutrino flux at Earth
after mixing and decohering,
and consequent relative strength of Glashow resonance, for six astrophysical models.
(Neutrinos and antineutrinos are shown separately, when they differ.) 
}
\label{table:nuebar}
\centering
\begin{tabular}{||c|c|c|c|c|c||}
\hline \hline 
 & \multicolumn{2}{c|}{} & \multicolumn{2}{c|}{} & \\
 &\multicolumn{2}{c|}{\ Source flavor ratio\ \ }  & \multicolumn{2}{c|}{\ Earthly flavor ratio\ \ }   &\ $\nuebar$ fraction in flux ($\CR$)\ \  \\ 
 					  & \multicolumn{2}{c|}{}     & \multicolumn{2}{c|}{} & \\ \hline 
 					  & \multicolumn{2}{c|}{}               & \multicolumn{2}{c|}{}           & \\ 
\ $pp\rarr\pi^\pm$~pairs\ \ & \multicolumn{2}{c|}{(1:2:0)} & \multicolumn{2}{c|}{(1:1:1)} & $18/108=0.17$   \\ 
 					  & \multicolumn{2}{c|}{}               & \multicolumn{2}{c|}{}            & \\ \hline   
  					  & \multicolumn{2}{c|}{} 		      & \multicolumn{2}{c|}{} & \\ 
\ \ w/ damped $\mu^\pm$  & \multicolumn{2}{c|}{(0:1:0)} & \multicolumn{2}{c|}{(4:7:7)} &  $12/108=0.11$  \\ 
 & \multicolumn{2}{c|}{} & \multicolumn{2}{c|}{} & \\ \hline 
      & & & & & \\ 
\ $p\gamma\rarr\pi^+$ only\ \ & \ (1:1:0)\ \ & (0:1:0) & (14:11:11) & (4:7:7) & $8/108=0.074$ \\ 
    & & & & & \\ \hline
     & & & & & \\ 
\ \ \ w/ damped $\mu^+$ & \ (0:1:0)\ \ & (0:0:0) & (4:7:7) & (0:0:0) &  0  \\ 
     & & & & & \\ \hline
 & \multicolumn{2}{c|}{} & \multicolumn{2}{c|}{} & \\ 
charm decay &  \multicolumn{2}{c|}{(1:1:0)} & \multicolumn{2}{c|}{(14:11:11)} & $21/108=0.19$  \\ 
 & \multicolumn{2}{c|}{} & \multicolumn{2}{c|}{} & \\ \hline 
      & & & & & \\ 
\  neutron decay\ \  & \ (0:0:0)\ \ & (1:0:0) &\ (0:0:0) & (5:2:2) & $60/108=0.56$    \\ 
 & & & & & \\ \hline\hline
\end{tabular}
\end{table}

When the $\pi^\pm$ mode occurs in an astrophysical source, 
isospin invariance yields a roughly equal ratio of $\pi^+$, $\pi^-$, and $\pi^0$ production,
followed by decay of the charged $\pi^\pm$s through the  $\mu^\pm$ chain to produce 
equal numbers of $\numu$ and $\numubar$, a number of $\nue$ plus $\nuebar$ equal to 
a half of $\numu$ plus $\numubar$, and roughly equal numbers of $\nue$ and $\nuebar$.
The rest-frame lifetimes of the charged pions and muons are $2.6\times 10^{-8}$~s and $2.2\times 10^{-6}$~s, respectively.
Since the rest frame lifetime of the muon exceeds that of the charged pion by a factor of 85, 
it is possible for $\pi^\pm$~decay to take place but the subsequent $\mu^\pm$ decay to be inhibited~\cite{dampedMU}. 
This would happen if the muon in the decay chain loses energy in the source environment before it decays 
(e.g., by synchrotron radiation in a $\vec{B}$-field, or by scattering). In a falling spectrum, the
decay of a lower-energy muon would make a negligible contribution.
This damped $\mu^\pm$ mode results in only $\numu$ and $\numubar$ being produced at the source;
flavor mixing between the source and Earth then produces a small amount of $\nuebar$.

In contrast to charged-pion production by $pp$~scattering, charged pions may be produced by $p\gamma$ scattering.
Here, the $\Delta^+$ resonance contributes to produce $\pi^+ +n$ and $\pi^0 +p$, in the ratio of $1:2$.  
Since $\pi^-$ production is suppressed and the  $\pi^+$ mode produces no $\nuebar$s at the source, 
only a small amount of $\nuebar$ arises from mixing~\cite{pgamma}.
If, in addition, the $\mu^+$s in $p\gamma$ mode are damped, then no antineutrinos are produced 
at all at the source, and so even with mixing there will be no $\nuebar$s at Earth. 

Charmed particles decay promptly (e.g. the $D^\pm$ has a lifetime of $1.0\times 10^{12}$~s) 
and semileptonically to $e^\pm$ or $\mu^\pm$ (e.g., the $D^\pm$ has a 34\% branching ratio to these modes). 
Lepton universality ensures that equal numbers (modulo small mass differences) of $\nue$, $\nuebar$, $\numu$, and $\numubar$ 
are produced, while production of $\nutau$ and $\nutaubar$ is kinematically suppressed.  
Thus, $\nuebar$s produced in charm decay will arrive at Earth.

Finally, there may be sources that inject a nearly pure neutron flux~\cite{betabeam}.
Such would be the case if Fe is emitted and subsequently dissociated to protons and neutrons, 
with the charged protons then degraded in energy, or swept aside, by a magnetic field at the source.
Such would also be the case if the cosmic accelerator entrains and accelerates charged protons,
with cosmic-ray escape occurring via $p_{\rm entrained} \rarr n+\pi^+$.  
This escaping (and pointing) beta beam decays to pure $\nuebar$, leading to a large amount of $\nuebar$ 
arriving at Earth, even after mixing.

Each of these six models are possible, as are combinations of the six.
For our purposes, we consider each model in isolation, and show how the rate for Glashow resonant events 
can serve as a barometer (``resonometer") distinguishing among these six source models.
A caveat is in order here.
It has been shown, especially in Ref.~\cite{Winter2010}, that multi-pion contributions can produce antineutrinos 
which via mixing ensure some $\nuebar$s at Earth.  These multi-pion contributions are not included in
our discussion here.  
For certain source parameters, the ``contamination'' from multi-pion processes can be large.
In addition, we assume that possible damping of muons at the sources is complete;
it may be incomplete, in which case results will be intermediate between the cases considered here. We mention in passing that the
effect of kaon decays on source neutrino flavor ratios is small in the energy range of interest~\cite{Winter2010}.
All in all, our results must be treated as suggestive. 
If and when Glashow resonance events are observed, a more careful treatment than presented here will be warranted.
Until  Glashow resonance events are observed, our results can be considered motivational.

At this early stage of astrophysical data collection, it is a good approximation~\cite{FuHoWei} to  assume that tribimaximal mixing~\cite{TBM} holds. 
Then, the evolution $\nua\rarr\nub$, with $\alpha$ and $\beta$ any elements of the three-flavor set
$\{e,\ \mu,\ \tau\}$, 
is described in terms of the PMNS matrix $U$, by the symmetric propagation matrix $P$ whose positive definite elements are
\beq{Ptheta13}
P_{\alpha\beta}= \sum_j  | U_{\alpha j} |^2\, | U_{\beta j} |^2=
\frac{1}{18}
\left(
\ba{rrr}
 10 & 4 & 4 \\
  4 & 7 & 7 \\
  4 & 7 & 7 \\
\ea
\right)
 \,.
 \eeq
The element with the largest uncertainty  is $P_{e\mu}$, which has an uncertainty of 20\% at 2$\sigma$. 
% This uncertainty will decrease with further experimentation.
From the last column of 
Table~\ref{table:nuebar} we see that all but one of the fifteen combinations of the six flux models predict a difference much larger than 20\% for the $\nuebar$ fraction.

\section{Resonant and Non-resonant events}
\label{sec:Breit-Wigner}
The resonant cross section for $\nuebar+e^-\rarr W^-\rarr {\rm hadrons}$ is 
\beq{Breit-Wigner}
\sigma_{\rm Res}(s) = 24\pi\,\Gamma_W^2\;{\rm B}(W^-\rarr\nuebar e^-)\,{\rm B}(W^-\rarr{\rm had})
\frac{(s/M_W^2)}{(s-M_W^2)^2 +(M_W \Gamma_W)^2}\,,
\eeq
where $M_W$ is the $W$ mass (80.4~GeV), $\Gamma_W$ is the $W$'s FWHM (2.1~GeV),
% $K$ is shorthand for,
and ${\rm B}(W^-\rarr\nuebar e^-)$ and ${\rm B}(W^-\rarr{\rm had)}$ are $W^-$ branching ratios
to the $\nuebar e^-$ state (11\%) and the hadronic state (67\%), respectively.  
At the peak, 
\beq{BWpeak}
\sigma_{\rm Res}^{\rm peak}(s) = \frac{ 24\pi\,{\rm B}(W^-\rarr\nuebar e^-)\,{\rm B}(W^-\rarr{\rm had}) }{M_W^2}
= 3.4\times 10^{-31}{\rm cm}^2\,.
\eeq
%
% Numerically, $K=7.8 {\rm GeV}^2$.
\\
Consequently, the resonant cross section may be written as 
\beq{BW2}
\sigma_{\rm Res} = \left[ \frac{\Gamma_W^2\,s}{(s-M_W^2)^2 +(M_W \Gamma_W)^2} \right]\,\sigma_{\rm Res}^{\rm peak}(s)\,. 
\eeq
%AND CROSS SECTION NEEDS TO BE CHECKED, FACTORS OF 2, ETC! \\
%BARGER AND PHILLIPS HELPS, BUT DOES NOT GIVE THE BW FOR NEUTRINO MODES; \\
%THE SPIN COUNTING MAY BE DIFFERENT.
\\
The $W$'s width is small compared to the $W$'s mass ($\frac{\Gamma_W}{M_W}=2.6\%$), 
and the experimental resolution will always exceed by far the $W$ width.
Thus, we are justified in using the ``narrow width approximation'' (NWA) throughout.
A contour integration in $s$ over the $s$-dependent bracketed expression in Eq.~\rf{BW2}, 
and the residue theorem, yields the value $\pi\,M_W\Gamma_W$.
Thus, the resulting NWA  is simply
\beq{NWA}
% \label{NWA}
\sigma(s)_{\rm Res} 
% &=& 24\pi^2\,{\rm B}(W^-\rarr\nuebar e^-)\,{\rm B}(W^-\rarr{\rm had})\,
%	\left(\frac{\Gamma_W}{M_W}\right)\,\delta(s-M_W^2)\,,   \nonumber \\
%	&=& 
  = (\pi M_W \Gamma_W)\,\sigma_{\rm Res}^{\rm peak}(s)\,\delta(s-M_W^2)\,,  
\eeq
and the number of resonant events per unit time and unit steradian is 
\bea
\label{ResRate}
\left(\frac{N}{T\Omega}\right)_{\rm Res} 
&=& 
N_e\,(\pi M_W\Gamma_W)\;\sigma_{\rm Res}^{\rm peak} 
% \,{\rm B}(W^-\rarr\nuebar e^-)\,{\rm B}(W^-\rarr{\rm had})\,\left(\frac{\Gamma_W}{M_W}\right)\,
% \left( \frac{K\,\pi}{M_W\,\Gamma_W} \right) 
	\int dE_{\nuebar}\,\left(\frac{dF_{\nuebar}}{dE_{\nuebar}}\right)\,\delta(s-M_W^2) \nonumber \\
	 & & \nonumber \\
	 & & \nonumber \\ 
 &=& 
\frac{N_p}{2m_e}\,(\pi M_W\Gamma_W)\,\sigma_{\rm Res}^{\rm peak} \,
\left. 
	\frac{dF_{\nuebar}}{dE_{\nuebar}} 
\right|_{E_{\nuebar}=6.3{\rm PeV}}\,,
\eea
where $N_e=N_p$ is the number of electrons or protons in the detector volume.

\begin{figure}
\centering
\includegraphics[height=0.60\linewidth]{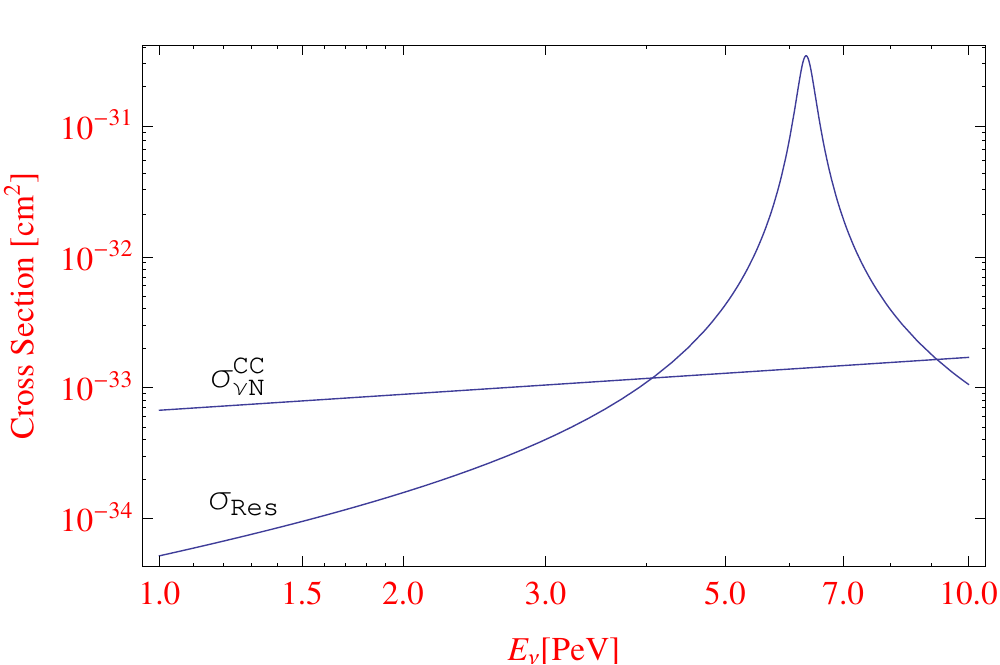}
\caption{Cross sections for the resonant process, $\nuebar+e^-\rarr W^-\rarr{\rm hadrons}$, and 
the non-resonant process, $\nue+N\rarr e^- +{\rm hadrons}$, in the 1--10~PeV region.}
\label{cross}
\end{figure}

% The effective area at the resonant energy is about 40x that of off-resonance events at a PeV 
% for the ``$pp$ mode'' neutrino/antineutrino flux (CHECK THIS). 
% This factor of 40 ($=6.3^{2.0}$) effectively cancels the falling power law ($E_\nu^{-\alpha}$) 
% of the incident neutrino spectrum~\cite{Ahlers:2005sn,Ema:2013nda}.
% Therefore, the expected event number at IceCube for a neutrino flux on-resonance at $\sim 6.3$~PeV, 
% relative to the three observed events just above a~PeV, 
% is  $3\times 40\times 6.3^{-\alpha+0.40}= 3\times (6.3)^{(2.40-\alpha)}$.
%
%
In contrast, the continuum (non-resonant) neutrino event rate 
% per flavor and per $\nu$ versus $\nubar$ 
between $\Emin\sim$~PeV to $\Emax$ is given by 
\beqa{nonResRate}
\left(\frac{N}{T\Omega}\right)_{\rm non-Res} \!\!\!
&=&  N_{n+p}\int_{\Emin}^{\Emax} dE_{\nu}\,\left(\frac{dF_{\nu}}{dE_{\nu}}\right)\,\sigma^{\rm CC}_{\nu N}(E_\nu) \nonumber 
\\ 
&\approx& \left.  
\frac{N_{n+p}}{(\alpha-1.40)} 
\left[ \left(\sigma_{\nu N}^{\rm CC}\,E_\nu\left(\frac{dF_\nu}{dE_\nu}\right)\right)\right|_{\Emin} \!\!\!
- \left. \left(\sigma_{\nu N}^{\rm CC}\,E_\nu\left(\frac{dF_\nu}{dE_\nu}\right)\right)\right|_{\Emax} \right ] \nonumber \\
&=& \left.    
\frac{N_{n+p}}{(\alpha-1.40)} 
% (6.3\,{\rm PeV})^{\alpha-1.40}
\left[ \left(\frac{6.3\,{\rm PeV}}{\Emin}\right)^{(\alpha-1.40)} \!\!\! -\left(\frac{6.3\,{\rm PeV}}{\Emax}\right)^{(\alpha-1.40)} \right]
% \left(\frac{6.3\,{\rm PeV}}{E_\nu^{\min}}\right)^{\alpha-1.40}
\left(\sigma_{\nu N}^{\rm CC}(E_\nu) \frac{E_\nu\,dF_\nu}{dE_\nu}\right)\right|_{E_\nu =6.3\,{\rm PeV}}  \,,
\eeqa
where $N_{n+p}$ is the number of nucleons in the detector volume, and 
$\frac{dF_\nu}{dE_\nu}$ is the total (summed over flavors) $\nu$ plus $\nubar$ flux.
Here we have assumed an $E^{0.40}$ energy dependence for $\sigma_{\nu N}$  
as predicted for the 1--10~PeV region in Ref.~\cite{GQRS},
and we have included only the charged-current cross section; 
in a falling spectrum, the neutral-current contribution is lower in average by 
$\frac{\sigma^{NC}_{\nu N}(E_{\rm obs}) }{\sigma^{CC}_{\nu N}(E_{\rm obs}) } \ymean^{\alpha-0.4}$,
where $\ymean=\frac{E_{\rm obs} }{E}$ is the average fraction of energy transferred from the incident neutrino to the detector.
The simple Fermi shock-acceleration mechanism yields $\alpha =2.0$,
whereas an earlier statistical study of the first-release dataset concluded that $\alpha$ was constrained by the
absence of Glashow events in the IceCube data to $\alpha\ge 2.3$~\cite{Anchordoqui:2013qsi,Anchordoqui:2013dnh}.
Taking $\ymean\sim 0.25$ and the NC to CC ratio to be 0.4, one finds less than a 5\% contribution from the neutral-current even with 
the conservative spectral index of $\alpha=2$.
The resonant cross section and the non-resonant charged-current $\sigma_{{\nue}_N}$ cross section are shown in Fig.~\ref{cross}.

From Eq.~\rf{nonResRate}, it is seen that the integrated continuum event rate scales with the minimum energy as
\beq{nonRes-scales}
\left(\frac{N}{T\Omega}\right)_{\rm non-Res} \propto \left[ \, ({\Emin})^{-(\alpha-1.40)} - ({\Emax})^{-(\alpha-1.40)}\,  \right] \,.
\eeq
Failure of future events to follow this energy-dependent rate equation would indicate a broken power-laaw spectrum, or in the extreme case, a cutoff spectrum.
On the other hand, when $\Emax$ can be taken to infinity, as can be done when the neutrino energy spectrum is a power-law falling as fast or faster
than $E^{-2}$, then we count all events that are initiated in the IceCube detector with energy exceeding $\Emin$.

We normalize the expected number of events in any energy interval to the expected number $\Nexp$ for the highest-energy IceCube 
bin with nonzero number, the 1-2~PeV bin.
Then, in the limit $\Emax\to \infty$, we have that the expected number of continuum events above $E_\nu^{\min}$ is
\beq{continuum}
N^{\rm expect} (\ge E_\nu) = \left( \frac{ E_\nu^{-(\alpha-1.40)} }{ 1-2^{-(\alpha-1.40)} }\right) \, \Nexp \,,
\eeq
which for $\alpha=2.0$ and~2.3 is equal to 
$2.94\,E^{-0.6}\,\Nexp$ and $2.15\,E^{-0.9}\,\Nexp$, respectively.
In turn, the number expected above 1~PeV is $2.94\,\Nexp$ and $2.15\,\Nexp$, respectively;
the number of events expected above 2~PeV is $1.94\,\Nexp$ and $1.15\,\Nexp$, respectively.

The 1-2~PeV IceCube bin contains the three observed events.
The expected event number for this bin is not known.
The ``Feldman-Cousins''~\cite{FeldCous} tables provide an estimate for the range of expected numbers of events,
given an observed number of events, with or without background.
(The zero-background case is the relevant one for us.)
Given three events in the 1-2~PeV bin, the Feldman-Cousins expected number of events for this bin 
is 0.82 to 8.25 at 95\%~C.L.
% (corresponding to $2\sigma$ if the distribution is Gaussian).
However, there is additional information in the IceCube data: no events are observed above $\sim 2$~PeV.
Thus, tension between the populated bin and the remaining unpopulated bins 
is minimized by investigating the lower numbers of expected events.
Consider the integer expected values $\nmean=$~1, 2, and 3 as representative;
the mean value 3 is appropriate if the observed value were spot on the mean, while the mean values 
1 and 2 are appropriate if the observed value is an upward fluctuation.
The Poisson probability to observe $n$ events against an expected number $\nmean$ is 
$P(n | \nmean)= e^{-\nmean} \, \frac{\nmean^n}{n!} $.
Thus, we have probabilities $P(3 | 3) = 22\%$ for the ``spot-on'' rate, and 
$P(3 | 2) = 18\%$ and $P(3 | 1) = 6.1\%$ for possible upward fluctuations.
Since we discount the disfavored cases where $\nmean > n$, we do not have the general Poisson result that 
$\int_0^\infty d\nmean\,P(n | \nmean) = 1$.
Thus, it is the relative rates 1, 0.82, and~0.28 for the expected values 3,~2,~1, respectively, that lead us to the obvious conclusion:
with just three events, the unknown expected number spans a large range of possibilities and so is ill-determined.

For $\Nexp =3,\,2,\,1$, we expect  $N(\ge 2~{\rm PeV})=5.8\,(3.5),\, 3.9\,(2.3),\, 1.9\,(1.15)$ events, for $\alpha=2.0$ (2.3), respectively.
No events above $\sim2$~PeV are observed.
The Poisson probability for a downward fluctuation to no events in a bin where $\nmean$ are expected is $P(0 | \nmean)=e^{-\nmean}$.
Thus, the tension between observed events in the 1-2~PeV bin and the absence of events above 2~PeV is quantified in 
the probabilities to observe none of the expected continuum events above 2~PeV:
0.30\%\,(3.0\%), 2.0\%\,(10\%), and 15\%\,(33\%), respectively. 
Moreover, if one normalizes to the three observed events not in the 1-2~PeV interval,
but rather in the 1-3~PeV interval, then the expected number of continuum events above 3~PeV is reduced to 3.2 (1.8),
with Poisson probabilities to observe no events of 4.1\%~(17\%).
As discussed above, these odds are higher if the three observed events are themselves an upward fluctuation.

At face value, these results favor the more steeply falling spectrum, 
and may even suggest a broken power law or cutoff~\cite{ABGLMPPW} in the neutrino spectrum.
However, these results are not compelling at present.

Here we will assume that the absence of events is the result of a downward fluctuation, and continue 
the calculation with the unbroken power spectrum to assess possibilities for the Glashow resonance event rate.
Since the event rate expected for the continuum and Glashow resonance depends on the expected rate determined with
$\sim$PeV events, one cannot yet predict the number of expected events at higher energy.
Nevertheless, in the ratio of expected Glashow events to expected continuum events, which we next present,  
the normalizing factor cancels out.

%\subsection{Ratio of Resonant to Non-Resonant Events}
%\label{subsec:eventratio}
%
From Eqs.~\rf{ResRate} and~\rf{nonResRate}, we find the ratio of resonant Glashow events to non-resonant continuum events to be
\beqa{RateRatio}
&&\frac{N_{\rm Res}}{N_{\rm non-Res}(E_\nu>E_\nu^{\min})} \nonumber \\
% && \ = 12\pi^2 \left(\frac{\Gamma_W}{M_W}\right) \,
% {\rm B}(W^-\rarr\nuebar e^-)\,{\rm B}(W^-\rarr{\rm had})\,
% \frac{(\alpha-1.40)\,\CR}{M_W^2\,\sigma_{\nu N}^{\rm CC}(E_\nu=6.3\,{\rm PeV)}}\,
% \left( \frac{E_\nu^{\min}}{6.3\,{\rm PeV}} \right)^{\alpha-1.40} \\
 && = \frac{10\,\pi}{18} \left(\frac{\Gamma_W}{M_W}\right) \left(\frac{\sigma_{\rm Res}^{\rm peak}}{\sigma_{\nu N}^{\rm CC}(E_\nu=6.3\,{\rm PeV)}}\right)
\frac{ (\alpha-1.40) \left( \frac{E_\nu^{\min}}{6.3\,{\rm PeV}} \right)^{\alpha-1.40} }{ \left[ 1-\left(\frac{\Emin}{\Emax}\right)^{(\alpha-1.40)} \right] }
\,\,\CR\,, \\
 && \nonumber \\
&& \ = 11\times \frac {(\alpha-1.40) \left( \frac{E_\nu^{\min}}{6.3\,{\rm PeV}} \right)^{\alpha-1.40}} {{ \left[ 1-\left(\frac{\Emin}{\Emax}\right)^{(\alpha-1.40)} \right] }}\times \CR\,,
{\rm \ \ \ with\ }
\CR\equiv 
\left[
	 { \left( \frac{dF_{\nuebar}}{dE_{\nuebar}} \right)}  /
% \right|_{E_{\nuebar}=6.3\,{\rm PeV} }
        { \left( \frac{dF_\nu}{dE_\nu}                   \right)} 
\right]_{E=6.3\,{\rm PeV} }
\nonumber\,.
\eeqa
Here we have taken $\frac{N_p}{N_p+N_n} = \frac{10}{18}$ in the detector material (water), and set 
$\sigma^{\rm CC}_{\nu N}=\sigma^{\rm CC}_{\nuebar N} = 1.42\times 10^{-33}~{\rm cm}^2$ at $E_\nu=6.3$~PeV~\cite{GQRS}.
$\CR$ is the ratio of the $\nuebar$ flux that produces the resonance events to the total $\nu$ flux
that produces the continuum events; $\CR$ is a model-dependent number, exhibited for each of our six models 
in the final column of Table~\ref{table:nuebar}. 
We stress that the ratio in Eq.~\rf{RateRatio} is valid for down-coming events, 
but not for up-coming events.  
The reason is that the large resonant cross section at 6.3~PeV implies that  
6.3~PeV neutrinos are strongly absorbed if transiting the Earth,
thereby eliminating the possibility for up-coming Glashow events~\cite{GQRS}. 
%In our estimates, we do not include neutrino absorption, and consider only down-coming events.
%The event ratio is now given by Eq.~\rf{RateRatio} with the denominator set equal to unity.

We list in Table~\ref{table:res2nonres} the ratio of Glashow events to continuum events above 
$E_\nu^{\min} = 1,\ 2,\ 3,\ 4,\ 5$~PeV, with $\alpha=2.0\ (2.3)$ and $\Emax=\infty$,
for the six models of cosmic neutrino production under consideration.{\footnote{
The purpose of allowing for a finite $\Emax$ in Eq.~\rf{RateRatio} is to compare our ratios 
to the ratios that result from the effective areas provided in Ref.~\cite{Aartsen:2013jdh}.
%and numerically in the ``Supplementary Material'' which accompanies the {\it Science} article. 
There an $\Emax=10$~PeV. 
On including this $\Emax$ in our calculation, 
we find very good agreement with the IceCube numbers.
Note that in the IceCube nomenclature for incident $\nu+\nubar$ fluxes, 
the ratio of down-coming Glashow events to continuum events is given by 
$\, \big(\frac{{\rm \nue} - {\rm \numu}}{3  {\rm \numu}}\big)_{south}$.}
Note that we keep the value of $E_\nu^{\min}$ well below the energy region of the resonance:
at the energy value of the peak minus one FWHM, the incident neutrino energy is 
$6.3\,{\rm PeV}\,(1-\Gamma_W/M_W)^2\approx 6.3\,{\rm PeV}\,(1-0.052)=6.0\,{\rm PeV}$.

We note that the numbers of expected resonant events presented in Table~\ref{table:res2nonres}
 is greatly reduced from the ratio of resonant to non-resonant cross sections by the additional factors.  
 The cross section ratio at 6.3~PeV is 240: see Fig.~\ref{cross}. The $\frac{\Gamma_W}{M_W}$ ratio is 1/38.
The $\alpha$-dependent factor $\left[ (\alpha-1.40) \left( \frac{1\,{\rm PeV}}{6.3\,{\rm PeV}} \right)^{\alpha-1.40}\right]$ 
yields about 0.2 for both $\alpha$'s of interest, 2.0 and 2.3.
The end result is about $2\CR$ for the ratio of resonant events to non-resonant events above 1~PeV.
% More precise numbers are listed in the table.

Of course, the expected number of Glashow events does depend on $\Nexp$.
The number of Glashow events is found by multiplying the first numerical column of 
Table~\ref{table:res2nonres} by $N(\ge 1\,{\rm PeV})=2.94\,\Nexp\,(2.15\,\Nexp)$.  
These expected resonant event numbers are 1.1 (0.69), 0.71 (0.43), 0.47 (0.28), 0 (0), 1.2 (0.77), and 3.5 (2.1),
each times~$\Nexp$, 
for the six models, and for $\alpha=2.0\,(2.3)$.
With increased statistics the Glashow event numbers may separate into values which discriminate among the astrophysical source models.

Since no 6.3~PeV events are observed, the Poisson probabilities for each model, based solely on the absence of resonance events, 
for $\Nexp=3$, are, 
3.8\% (13\%), 12\% (28\%), 24\% (43\%), large (large), 2.7\% (9.9\%), and 0.0025\% (0.18\%), respectively;
and for $\Nexp=1$, are
34\% (50\%), 49\% (65\%), 62\% (76\%), large (large), 30\% (46\%), 3.0\% (12\%), respectively.
All models remain viable except perhaps the final one,
where neutron decay to pure $\nuebar$ predicts some resonance events at Earth.
However, since the probabilities vary exponentially with $\Nexp$, more data is needed before 
reasonably-definite conclusions can be drawn.
%\footnote
%{We note that simple fits to the shower/track ratio of the data~\cite{Sergio} favor a flux with 
%enhanced $\nue+\nuebar$, a result seemingly in serious tension with the absence of resonance events.
%}
% The six models yield Poissonian occurrence probabilities with $\Nexp=????bogus$ of 
% 37\%, 50\%, 67\%, 100\%, 33\%, and 4\% for $\alpha=2$, and slightly larger probabilities for $\alpha=2.3$.
These ``Glashow-event'' probabilities should be multipled by the continuum probabilities to determine overall Poisson probabilities
for a $\Nexp$ value, and for the unbroken power law hypotheses with $\alpha=2.0$~and 2.3.
\begin{table}[t]
\caption
{Ratio of resonant event rate around the 6.3~PeV peak  
to non-resonant event rate above $E_\nu^{\min}=1,\ 2,\ 3,\ 4,\ 5$~PeV. 
The single power-law spectral index $\alpha$ is taken to be 2.0 and  2.3 for the non-parenthetic 
and parenthetic values, respectively.
The single power-law extrapolation just above 1~PeV predicts 
a mean number of observed resonance events around 6.3~PeV equal to the first numerical column 
times $2.94\,\Nexp\,\,(\,2.15\,\Nexp\,)$, as calculated in the text.
}
\label{table:res2nonres}
 \begin{tabular}{||c|c|c|c|c|c||}
\hline \hline 
      & & & & &  \\
 $E_\nu^{\min}$ (PeV) & 1 & 2 & 3 & 4 & 5 \\
      & & & & & \\ \hline \hline
      & & & & & \\
\ \ $pp\rarr \pi^\pm$~pairs \  & \ 0.37 \ (0.32) \   & \ 0.56 \ (0.59) \  & \ 0.71 \ (0.85) \ & \ 0.84 \ (1.1) \    & \ 0.96 \ (1.3) \  \\ 
      & & & & & \\ \hline  
      & & & & & \\
w/ damped $\mu^\pm$        & \ 0.24 \ (0.20) \   & \ 0.37 \ (0.38 ) \ & \ 0.47 \ (0.56) \ &  \ 0.54 \ (0.71) \ & \ 0.62 \ (0.88) \  \\ 
      & & & & &  \\  \hline
      & & & & & \\ 
\ \ $p\gamma\rarr\pi^+$ only \ & \ 0.16 \ (0.13) \ & \ 0.24 \ (0.26 )  \ & \ 0.31 \ (0.37) \ &  \ 0.37 \ (0.48)  \ &  \ 0.42 \ (0.59)\  \\ 
     & & & & & \\ \hline
     & & & & & \\ 
w/ damped $\mu^+$            & \ 0\ \ (0)	 \     	& \ 0\ \ (0) \        & \ 0 (0) \            & \ 0\ \ (0) \           &  \ 0\ \ (0)  \     \\ 
     & & & & & \\ \hline
     & & & & & \\ 
charm decay                     & \ 0.41 \ (0.36) \   & \ 0.62 \ (0.67) \   & \ 0.80 \ (0.95) \ & \ 0.94 \ (1.2)  \    &  \ 1.1 \ (1.6) \  \\ 
    & & & & & \\ \hline
    & & & & & \\ 
 neutron decay                 & \ 1.2 \ (1.0) \     & \ 1.9 \ (2.0) \      & \ 2.3 \ (2.8) \ & \ 2.8 \ (3.6)  \     & \ 3.2 \ (4.4) \     \\
 
 & & & & &  \\ \hline\hline
\end{tabular}
\end{table}
%
%\subsection{Exotic Alternatives to Suppress the Glashow Resonance}
%\label{Exotica}
%

For the sake of completeness, we briefly consider the possibility of exotic neutrino properties that modify the flavor mix of neutrinos, specifically neutrino decay and pseudo-Dirac neutrino oscillations.
Neutrino decay~\cite{NuDK}  allows the flavor mix to deviate significantly from the democratic mix. Observation of a significant $\nuebar$ flux from SN1987A precludes any observable effects 
of $\nu_1$ decay on $L/E$ scales of astrophysical interest. In the case of a normal hierarchy (with mass ordering $m_{\nu_1} < m_{\nu_2} < m_{\nu_3}$), the $\nu_2$ and $\nu_3$ mass eigenstates may decay completely to $\nu_1$, 
whose flavor content ratios are $|U_{e1}|^2:|U_{\mu 1}|^2:|U_{\tau 1}|^2=4:1:1$ for both $\nu$ and $\nubar$.
The $\nuebar$ content of the neutrino flux at Earth is then $1/3$ which may be an {\it enhancement}. 
On the other hand, if the mass hierarchy is inverted
(with $m_{\nu_3}<m_{\nu_1}<m_{\nu_2}$), then 
both $\nu_1$ and $\nu_3$ are stable and a variety of final flavor ratios are possible,
depending on the intial ratios of $\nu_1$, $\nu_2$, and $\nu_3$, and the decay mode of $\nu_2$.
% In neither case is the resonant event rate suppressed.
% /footnote
% {As an example, consider the popular initial ratio (1:2:0),
% which evolves to a ratio (1:1:1) in both the flavor and mass basis.
% Upon $\nu_2$ decay to $\cos^2\theta\,\nu_1 +\sin^2\theta\,\nu_3$,
% the flavor ratios at Earth are (4:4:4)$+\cos^2\theta$(4:1:1)$+\sin^2\theta$(0:3:3).
% For decay to a pure mode, we get (8:5:5) for decay to $\nu_1$, and (4:7:7) for decay to $\nu_3$;
% so depending on the angle $\theta$, any ratio intermediate to these two is possible.
% A positive remark is that, in the context of the decay model,
% an inference of the decay mode may be possible from the flavor ratios observed at Earth.}

Another possibility for deviations from standard flavor mixes~\cite{pDirac} arises in scenarios of 
pseudo-Dirac neutrinos~\cite{earlyPDirac}, 
in which each of the three neutrino mass eigenstates is a doublet with tiny mass differences 
less than $10^{-6}$~eV (to evade detection so far).\footnote
{
In fact, observing an energy-dependence of flavor mixes of high energy cosmic neutrinos 
is the only known way to detect mass-squared differences in the range 
$10^{-18}-10^{-12}$~eV$^2$.
}
The smallness of the mass difference tells us that the mixing angle 
between the active state with $SU(2)$~couplings, 
and the sterile state without, is necessarily maximal.
For cosmically-large $L/E$, the flux of each active flavor is therefore reduced by a half.
Of course, if all three flavors are reduced by a half, there is no change in the flavor ratios;
however, at intermediate energies each flavor can be reduced or not, leading to 
a possible suppression  of the absolute flavor ratio for $\nuebar$
by $R^{\rm pD}_{\nuebar}/R_{\nuebar}$ of roughly 1/2, or an enhancement of the $\nuebar$ flux ratio of roughly 2.
(Note that the maximal suppression/enhancement will be a bit less than 1/2 or 2 if there is a  $\nue$ flux present.)

\section{Conclusions}
\label{sec:conclusions}
Normalized to the three down-coming IceCube events in the 1-2~PeV range, 
we find that the number of predicted resonant Glashow events ranges from zero 
(for the damped $\mu^+$ mode, which generates no antineutrinos)
to almost three (for the neutron decay mode which generates only antineutrinos) times $\Nexp$.
The other four popular neutrino-generating modes give intermediate values.
Thus we have demonstrated that the fraction of resonance events is a potential discriminator among the 
popular neutrino-generating astrophysical models.

Our calculations are done in a somewhat idealized approximation.
For example, in pion production from $p\gamma$ collisions, 
we consider only the contribution of the $\Delta^+$ intermediate states.
Also, we do not consider the possibility that more than one neutrino source model may be contributing.
When more data become available, refinements on our ``Resonometer'' will become necessary.

Until that day, we conclude that the absence of Glashow resonance events in IceCube favors 
the lower values of the fractional $\nuebar$ flux.
Should this non-observation of resonance events continue, the ``damped $\mu^+$ mode'' 
$p\gamma\rarr \pi^+ n \rarr n+\mu^+  +\numu$ would become uniquely favored.\footnote
{
In Ref.~\cite{Anchordoqui:2014yva} it was noted that a damped $\mu^+$ mode will suppress antineutrinos 
and therefore the Glashow event rate, but will also generate in IceCube ``double-bang" events 
in the $3-10$~PeV range via $\numu$~oscillations to $\nutau$'s.
}
Caveats to this conclusion include the possibility of pseudo-Dirac neutrino oscillations,
and the possibility of neutrino decay.

%\newpage
%
\section*{Acknowledgments}
We thank Haim Goldberg for bringing the relevance of the Feldman-Cousins statistics to our attention.
% LAA is supported by U.S. National Science Foundation (NSF) CAREER Award PHY1053663 
% and by the National Aeronautics and Space Administration (NASA) Grant No. NNX13AH52G. 
VB is supported by the U. S. Department of Energy (DoE) Grant No. DE-FG-02- 95ER40896.  
% HG is supported by NSF Grant No. PHY-0757959.  JGL is supported by DoE Grant No. DE-FG02-04ER41291.  
LF is supported in part by US DoE grant DE-FG05-85ER40226.
JGL, DM and SP are supported by DoE Grant No. DE-SC0010504, 
and SP additionally by the Alexander Von Humboldt foundation.  
% TCP is supported by NSF Grant No. PHY-1205854 and NASA Grant No. NNX13AH52G. 
TJW is supported by DoE Grant No. DE-FG05-85ER40226 and the Simons Foundation Grant No. 306329.

\end{document}